\documentclass[aps,prd,twocolumn,superscriptaddress,tightenlines,nofootinbib]{revtex4-1}
\usepackage[T1]{fontenc}
\usepackage{microtype}
\usepackage[textsize=scriptsize,backgroundcolor=red!70,linecolor=red]{todonotes}
\usepackage{booktabs}
\usepackage{dcolumn}
\usepackage{amsmath}
\usepackage{amsfonts}
\usepackage{amssymb}
\usepackage{graphicx}
\usepackage{hyperref}
\usepackage[all]{hypcap}
\usepackage{paralist}
\usepackage{multirow}

\newcommand{\chisq}{\ensuremath{\chi^2}}

\usepackage{graphicx}
\usepackage{dcolumn}
\usepackage{bm}
\usepackage{epsf}
\usepackage{dcolumn}
\def\be{\begin{equation}}
\def\ee{\end{equation}}
\def\bea{\begin{eqnarray}}
\def\eea{\end{eqnarray}}
\def\gsim{\ \rlap{\raise 2pt\hbox{$>$}}{\lower 2pt \hbox{$\sim$}}\ }
\def\lsim{\ \rlap{\raise 2pt\hbox{$<$}}{\lower 2pt \hbox{$\sim$}}\ }
\def\dslash{\kern-4pt \not{\hbox{\kern-2pt $\partial$}}}
\def\pslash{\not{\hbox{\kern-2pt p}}}

\begin{document}
\DeclareGraphicsExtensions{.eps,.ps}


\title{Nonmaximal neutrino mixing at NO$\nu$A from nonstandard interactions}



\author{Jiajun Liao}
\affiliation{Department of Physics and Astronomy, University of Hawaii at Manoa, Honolulu, HI 96822, USA}
 
\author{Danny Marfatia}
\affiliation{Department of Physics and Astronomy, University of Hawaii at Manoa, Honolulu, HI 96822, USA}

\author{Kerry Whisnant}
\affiliation{Department of Physics and Astronomy, Iowa State University, Ames, IA 50011, USA}

\begin{abstract}

Muon neutrino disappearance measurements at NO$\nu$A suggest that maximal $\theta_{23}$ is excluded at the 2.6$\sigma$ CL. This is in mild tension with T2K data which prefer maximal mixing. Considering that NO$\nu$A has a much longer baseline than T2K, we point out that the apparent departure from maximal mixing in NO$\nu$A may be a consequence of nonstandard neutrino propagation in matter. 

\end{abstract}
\maketitle

Recently, NO$\nu$A released a new measurement of $\theta_{23}$ from the $\nu_\mu$ disappearance channel which indicates that $\theta_{23}=\pi/4$ is excluded at the 2.6$\sigma$ CL~\cite{NOvAdata}. T2K measurements in the same channel prefer $\theta_{23}=\pi/4$~\cite{Abe:2014ugx}. Neutrinos in both the NO$\nu$A and T2K experiments travel through
a long distance in matter, so that their oscillation probabilities are modified by the interactions with matter via the MSW effect~\cite{Wolfenstein:1977ue, Mikheev:1986gs}. Since the NO$\nu$A  baseline (810~km) and neutrino energy ($\sim$~2~GeV) are greater than those for T2K (295~km and $\sim$~0.6~GeV), the matter effect in the NO$\nu$A experiment is much larger than in T2K. However, the standard weak interactions with matter have a negligible effect on the $\nu_\mu$ survival probabilities. In this Letter, we study the matter effects induced by nonstandard interactions (NSI) on the $\nu_\mu$ survival probabilities, and show that they reconcile the discrepancy between the NO$\nu$A and T2K measurements of $\theta_{23}$.

NSI are motivated by physics beyond the standard model (SM), and provide a model-independent way to study subdominant effects in neutrino oscillation experiments; for recent reviews, see Ref.~\cite{Ohlsson:2012kf, Miranda:2015dra}. NSI can in general affect neutrino production, detection, and propagation in matter. Here we focus on the matter NSI, which can be described in an effective theory by the dimension-six operators~\cite{Wolfenstein:1977ue}
\be
  \label{eq:NSI}
  \mathcal{L}_\text{NSI} =-2\sqrt{2}G_F
   \epsilon^{fC}_{\alpha\beta} \!
        \left[ \overline{\nu_\alpha} \gamma^{\rho} P_L \nu_\beta \right] \!\!
        \left[ \bar{f} \gamma_{\rho} P_C f \right] + \text{h.c.}\,,
\ee
where $\alpha, \beta=e, \mu, \tau$, $C=L,R$, $f=u,d,e$, and
$\epsilon^{fC}_{\alpha\beta}$ are dimensionless parameters that quantify the
strength of the new interaction in units of the Fermi constant $G_F$. The Hamiltonian that describes neutrino propagation in matter with NSI is

\be
H = \frac{1}{2E} \left[ U\text{diag}(0,\Delta m^2_{21},\Delta m^2_{31})
U^\dagger + V\right]\,,
\ee 
where $U$ is the Pontecorvo-Maki-Nakagawa-Sakata mixing matrix~\cite{Agashe:2014kda}, $\Delta m^2_{ij}=m^2_i-m^2_j$, and 
\be
V = A \left(\begin{array}{ccc}
1 + \epsilon_{ee} & \epsilon_{e\mu} & \epsilon_{e\tau}
\\
\epsilon_{e\mu}^* & \epsilon_{\mu\mu} & \epsilon_{\mu\tau}
\\
\epsilon_{e\tau}^*& \epsilon_{\mu\tau}^* & \epsilon_{\tau\tau}
\end{array}\right)\,.
\ee
Here, $A=2\sqrt{2}G_FN_eE_\nu$, each $\epsilon_{\alpha\beta}\equiv\sum\limits_{f,C}\epsilon^{fC}_{\alpha\beta}\frac{N_f}{N_e}$ gives the {\it effective} strength of NSI relative to the SM charged-current interaction in matter, and $N_f$ is the number density of fermion $f$. 

The effects of matter NSI on neutrino oscillations at NO$\nu$A have been analyzed at the probability level in Ref.~\cite{Friedland:2012tq}. We recently showed that matter NSI could lead to wrong determinations of the Dirac CP phase, the mass hierarchy, and $\theta_{23}$ octant at NO$\nu$A and T2K~\cite{Liao:2016hsa}. For example, the current hint of $\delta_\text{CP}=-\pi/2$ from T2K~\cite{Abe} could be due to a nonzero $\epsilon_{e\tau}$~\cite{Forero:2016cmb}.  Previous analyses of NO$\nu$A have focused on the $\nu_e$ appearance channel, in which the NSI terms related to $\epsilon_{ee}$, $\epsilon_{e\mu}$ and $\epsilon_{e\tau}$ are dominant~\cite{Liao:2016hsa}. However, in the $\nu_\mu$ disappearance channel, the leading NSI contributions come from the $\mu-\tau$ sector.

The Super-Kamiokande experiment has obtained the strong 90\%~CL constraints, 
$|\epsilon_{\mu\tau}|<0.011$ and 
$|\epsilon_{\tau\tau}-\epsilon_{\mu\mu}|<0.049$~\cite{Mitsuka:2011ty}, using a two-flavor analysis of its atmospheric neutrino data. However, it has been shown that the two-flavor framework is not adequate to constrain NSI parameters using atmospheric neutrino experiments~\cite{Friedland:2004ah, Friedland:2005vy}. Note that the Super-Kamiokande collaboration also performed a three-flavor analysis in Ref.~\cite{Mitsuka:2011ty} with the standard oscillation parameters fixed. As shown in Ref.~\cite{Friedland:2004ah}, constraints on $\epsilon_{\tau\tau}-\epsilon_{\mu\mu}$ are significantly weaker when the standard oscillation parameters are marginalized over. This is confirmed in a global three-flavor analysis of neutrino oscillation data including matter NSI, which yields the approximate 3$\sigma$ CL bounds, (taking $\epsilon_{\alpha\beta}^e=0$, and $|\epsilon_{\alpha\beta}|  \lesssim {N_u \over N_e} 
|\epsilon_{\alpha\beta}^u|)$, where $\epsilon_{\alpha\beta}^f \equiv \epsilon_{\alpha\beta}^{fL}+ \epsilon_{\alpha\beta}^{fR}$ and $N_u/N_e \simeq N_d/N_e \approx 3$),
$|\epsilon_{\mu\tau}|\lesssim 0.10$, $|\epsilon_{e\tau}| \lesssim 1.34$ and $-0.68\lesssim\epsilon_{\tau\tau}-\epsilon_{\mu\mu}\lesssim 0.66$; these are the limits from the SNO-DATA variant of the solar analysis in Ref.~\cite{Gonzalez-Garcia:2013usa}.\footnote{If instead, we assume uncorrelated errors and take the separate bounds on 
$|\epsilon_{\alpha\beta}^u|$ and $|\epsilon_{\alpha\beta}^d|$
in quadrature, i.e., $|\epsilon_{\alpha\beta}|  \lesssim \sqrt{({N_u \over N_e} 
|\epsilon_{\alpha\beta}^u|)^2 + ({N_d \over N_e} |\epsilon_{\alpha\beta}^d|)^2}$, we obtain the more conservative 3$\sigma$~CL bounds, $|\epsilon_{\mu\tau}| \lesssim 0.14$, $|\epsilon_{e\tau}| \lesssim 1.86$ and $-0.94 \lesssim \epsilon_{\tau\tau}-\epsilon_{\mu\mu} \lesssim 0.91$.}
Since Ref.~\cite{Gonzalez-Garcia:2013usa} only considered NSI with one flavor $f=e$, $f=u$ or $f=d$ at a time in the analysis of solar data, we consider these bounds to be representative. 



{\bf Nonmaximal mixing from NSI.}
To understand the dependence of the survival probabilities on the NSI parameters at NO$\nu$A and T2K, we first consider the two-flavor framework. The Hamiltonian that describes nonstandard neutrino propagation in matter induced by NSI in the $\mu-\tau$ sector is
\bea
H&=&\frac{\Delta m_{32}^2}{2E_\nu}\left[\begin{pmatrix}
   s^2_{23} & s_{23}c_{23} \\
   s_{23}c_{23} & c^2_{23}
   \end{pmatrix}+ \hat{A} \begin{pmatrix}
    \epsilon_{\mu\mu} & \epsilon_{\mu\tau} \\
    \epsilon_{\mu\tau}^* & \epsilon_{\tau\tau} 
    \end{pmatrix} \right]\,,
\label{eq:H}
\eea
where $\hat{A}=\frac{A}{\Delta m_{32}^2}$, and $c_{ij}$ ($s_{ij}$) denotes $\cos\theta_{ij}$ ($\sin\theta_{ij}$).

The matter density is constant for the relevant baselines,
and the $\nu_\mu$ survival probability in the two-flavor framework can be written in the form~\cite{Kopp:2010qt} 
\be
P(\nu_\mu \rightarrow \nu_\mu)=1-\sin^2 2\theta \sin^2 \left(\frac{\Delta m^2 L}{4E_\nu}\right)\,,
\label{eq:prob}
\ee
where 
\bea
{\Delta m^2 \over \Delta m_{32}^2} &=& \sqrt{(\cos 2\theta_{23}+(\epsilon_{\tau\tau}-\epsilon_{\mu\mu})\hat{A})^2+|\sin 2\theta_{23}+2\epsilon_{\mu\tau}\hat{A}|^2}\,,
\nonumber \\
\sin^2 2\theta &=& \left(1+\frac{(\cos 2\theta_{23}+(\epsilon_{\tau\tau}-\epsilon_{\mu\mu})\hat{A})^2}{|\sin 2\theta_{23}+2\epsilon_{\mu\tau}\hat{A}|^2}\right)^{-1}\,.\nonumber
\label{eq:amp2}
\eea

As can be seen from the above equations, even with maximal mixing in vacuum, i.e., $\theta_{23}=\pi/4$, the NSI terms can generate nonmaximal mixing in matter. Also, because $\cos 2\theta_{23}\ll \sin 2\theta_{23}$, the diagonal parameter $\epsilon_{\tau\tau}-\epsilon_{\mu\mu}$ has a larger effect on the deviation from maximal mixing than the off-diagonal parameter $\epsilon_{\mu\tau}$. This, coupled with the fact that $\epsilon_{\mu\tau}$ is more tightly constrained than $\epsilon_{\tau\tau}-\epsilon_{\mu\mu}$, leads us to fix $\epsilon_{\mu\tau}=0$.
 
In the three-flavor framework, we ignore the solar mass-squared difference (since $\Delta m_{21}^2/|\Delta m_{32}^2|\approx 0.03$), take the NSI parameters to be real, and the CP phase to be vanishing. Henceforth, we set $\epsilon_{\mu\mu}=0$, as the oscillation probabilities are not affected by subtracting an overall diagonal term in the Hamiltonian. We only consider nonzero $\epsilon_{\tau\tau}$ and $\epsilon_{e\tau}$ for simplicity.\footnote{In anticipation of our numerical results, 
we mention that our 
NO$\nu$A data analysis is insensitive to the $\cal O$(1) values of $\epsilon_{ee}$ allowed by the global fit of Ref.~\cite{Gonzalez-Garcia:2013usa}. Consequently, our conclusions are unaffected by an $\cal O$(1) $\epsilon_{ee}$ invoked to satisfy the relation, $|\epsilon_{e\tau}|^2 \simeq \epsilon_{\tau\tau} (1+\epsilon_{ee})$, required by high energy atmospheric data~\cite{Friedland:2004ah}.} The Hamiltonian in matter becomes
\bea
H&=&\frac{\Delta m_{32}^2}{2E_\nu}R_{23}\left[\begin{pmatrix}
   s_{13}^2 & 0 & c_{13}s_{13} \\
   0 & 0 & 0 \\
   c_{13}s_{13} & 0 & c_{13}^2
   \end{pmatrix}\right.
   \nonumber \\
   &+& \left.
    \hat{A} R_{23}^T\begin{pmatrix}
    1 & 0 & \epsilon_{e\tau} \\
    0 & 0 & 0 \\
    \epsilon_{e\tau} & 0 & \epsilon_{\tau\tau} 
    \end{pmatrix} R_{23}\right]R_{23}^T\,,
\label{eq:H3}
\eea
where $R_{ij}$ is a real rotation by an angle $\theta_{ij}$ in the $ij$ plane. If we assume the terms in the square bracket of Eq.~(\ref{eq:H3}) are diagonalized by $U^\alpha=R_{23}^\alpha R_{13}^\alpha R_{12}^\alpha$, where $R_{ij}^\alpha$ is a real rotation by an angle $\alpha_{ij}$ in the $ij$ plane, then the mixing matrix that diagonalizes the Hamiltonian in matter is $U^m=R_{23}U^\alpha$. Since $\hat{A}\approx 0.17 \frac{E_\nu}{2\, \text{GeV}}$ at NO$\nu$A, we find~\cite{Liao:2015rma}
\bea
\label{eq:delta23}
\alpha_{23}&=&-\frac{s_{23}c_{23}\epsilon_{\tau\tau}\hat{A}}{c_{13}^2+\cos2\theta_{23}\epsilon_{\tau\tau}\hat{A}}+{\cal O}(s_{13}\hat{A})\,,
\\
\label{eq:delta13}
\alpha_{13}&=&\frac{c_{13}s_{13}+c_{23}\epsilon_{e\tau}\hat{A}}{\lambda-s_{13}^2-\hat{A}}+{\cal O}(s_{13}\hat{A})\,,
\eea
where 
\bea
\label{eq:lambda}
\lambda &=&\frac{1}{2}\left(c_{13}^2+\epsilon_{\tau\tau}\hat{A}+\right.
\nonumber\\
& &\left.\sqrt{c_{13}^4+\epsilon_{\tau\tau}^2\hat{A}^2+2\cos2\theta_{23}c_{13}^2\epsilon_{\tau\tau}\hat{A}}\right)\,.
\eea
Then the $\nu_\mu$ disappearance probability can be written in the form of Eq.~\ref{eq:prob}, with the oscillation amplitude replaced by
\be
\sin^2 2\theta=\cos^4\theta_{13}^m\sin^2 2\theta_{23}^m+\sin^2\theta^m_{23}\sin^2 2\theta_{13}^m\,,
\label{eq:amp3}
\ee
where $\theta^m_{23}=\theta_{23}+\alpha_{23}$ and $\theta^m_{13}=\alpha_{13}$.

{\bf Data analysis.}
To analyze NO$\nu$A's $\nu_\mu$ disappearance results we extract the unoscillated spectrum, backgrounds, and data from Ref.~\cite{NOvAdata}. Since the data above 2.5~GeV are noisy and have an insignificant effect on the parameter fit~\cite{NOvAdata}, we only include 7~bins in the energy range [0.75~GeV, 2.5~GeV] in our analysis. The expected number of events per bin $N_i^\text{th}$ is calculated as 
\be
N_i^\text{th}=N_i^\text{unosc}\times \left\langle P(\nu_\mu\rightarrow \nu_\mu)\right\rangle _i + N_i^\text{bkg},
\ee
where $N_i^\text{unosc}$ is the expected number of events without oscillations, $N_i^\text{bkg}$ is the expected background, and  $\left\langle P(\nu_\mu\rightarrow \nu_\mu)\right\rangle _i$ is the average survival probability in each bin. We calculate the survival probabilities in the three-flavor framework using the GLoBES software~\cite{GLOBES} supplemented with the results of Ref.~\cite{Kopp:2006wp}. We choose $\theta_{12}$, $\Delta m_{21}^2$ to be the global best-fit values~\cite{Agashe:2014kda}, vary $|\Delta m_{32}^2|$ between 
$(2.0 - 3.5)\times 10^{-3}$~$\text{eV}^2$
and set $\delta_\text{CP}=0$ because the CP phase has a negligible effect on our analysis. 

To evaluate the significance of each scenario, we define 
\bea
\chisq &=& \sum_{i=1}^{7}  \left(\frac{ N_i^\text{th} - N_i^\text{obs}}{\sigma_i}\right)^2 \,,
\eea
where $N_i^\text{obs}$ is the observed number of events in each bin, and $\sigma_i$ is obtained by summing the statistical and systematic uncertainties in quadrature.
Both sets of (asymmetric) uncertainties are extracted from Ref.~\cite{NOvAdata}.

{\bf Results.} 
For the SM we find that the best fit value of $\sin^2\theta_{23}$ is 0.41 in the first octant and 0.63 in the second octant. Defining 
\begin{equation}
\Delta \chi^2(\epsilon_{\tau\tau},\epsilon_{e\tau})=\chi^2(\epsilon_{\tau\tau},\epsilon_{e\tau},\theta_{23}=\pi/4)-\chi^2_\text{min}(\epsilon_{\tau\tau},\epsilon_{e\tau}) \nonumber \,,
\end{equation}
 we find that in the SM case ($\epsilon_{\tau\tau}=\epsilon_{e\tau}=0$) maximal mixing is excluded at the 2.2$\sigma$ CL for the normal hierarchy, which is close to the NO$\nu$A result of 
2.6$\sigma$~CL~\cite{NOvAdata}.  

\begin{figure}
\includegraphics[width=0.45\textwidth]{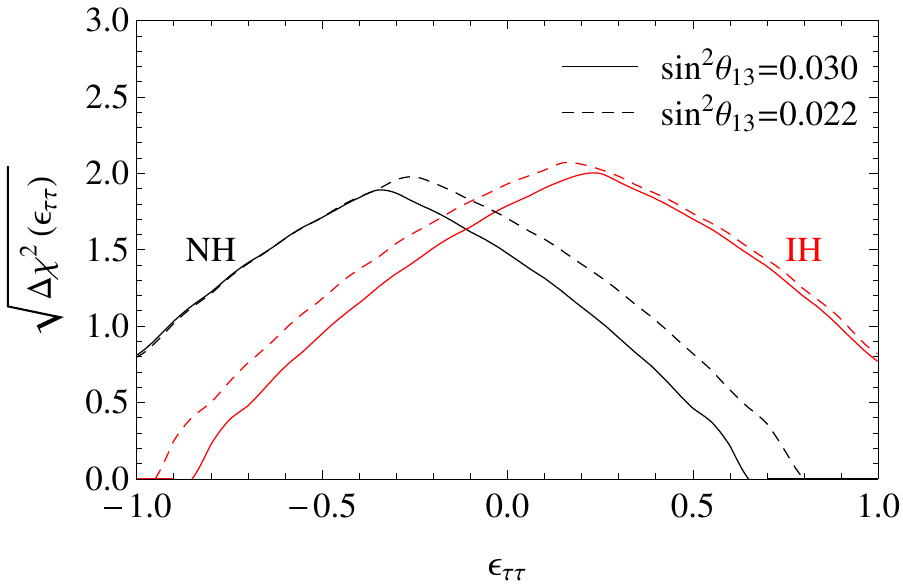}
\caption{The confidence level at which maximal mixing is excluded as a function of $\epsilon_{\tau\tau}$ with $\epsilon_{\mu\mu}=\epsilon_{\mu\tau}=0$ and \mbox{$|\epsilon_{e\tau}|< 1.2$}. The solid (dashed) curves correspond to $\sin^2\theta_{13}=0.030$ ($\sin^2 \theta_{13}=0.022$), and the black (red) curves correspond to the normal (inverted) hierarchy. }
\label{fig:chiett}
\end{figure}

\begin{figure*}[t]
	\centering
	\includegraphics[width=0.45\textwidth]{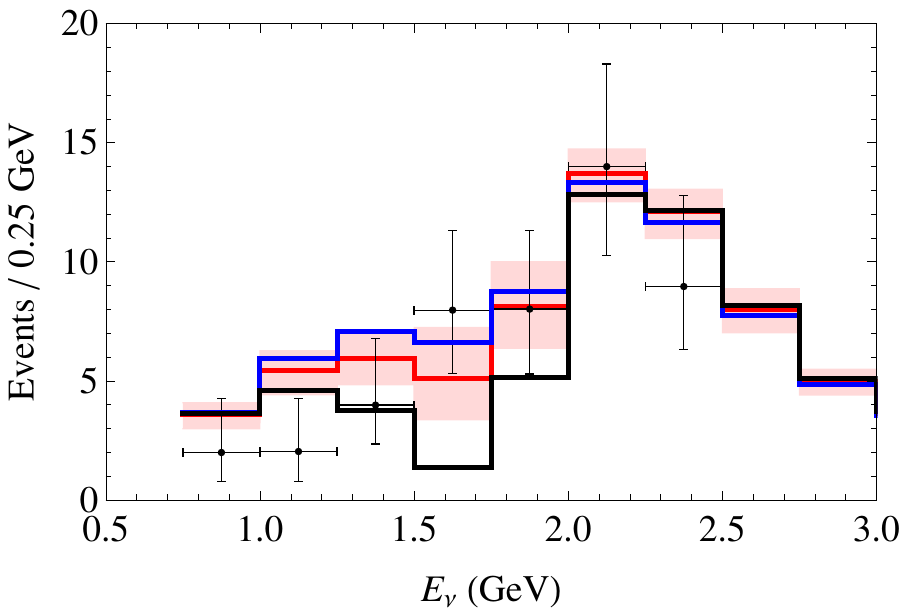}
	\includegraphics[width=0.45\textwidth]{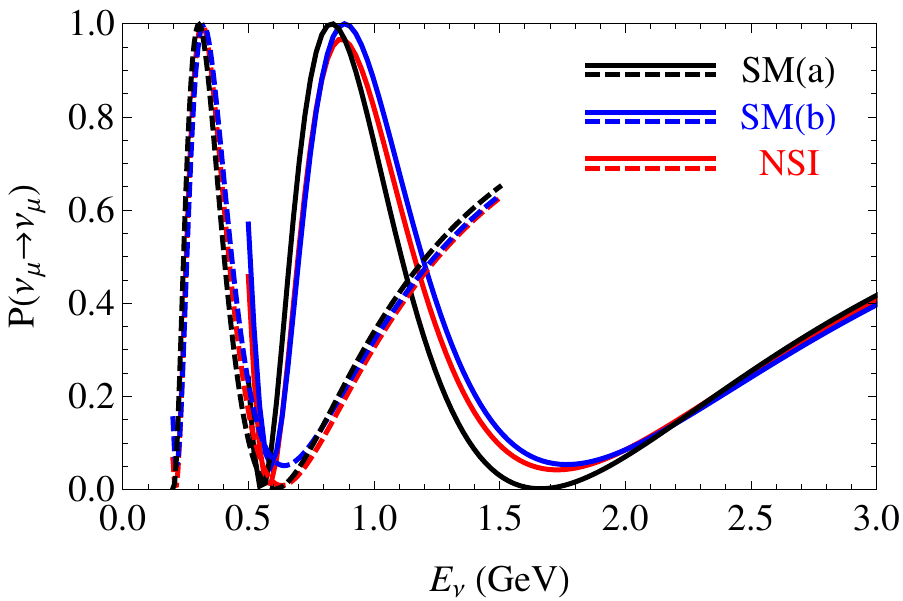}
	\caption{Comparison of the event distributions at NO$\nu$A and survival probabilities for three scenarios. SM(a): $\sin^2 \theta_{23} = 0.5$ and $\Delta m^2_{32}=2.51\times 10^{-3}$~eV$^2$; SM(b): $\sin^2 \theta_{23} = 0.4$ and $\Delta m^2_{32}=2.67\times 10^{-3}$~eV$^2$; NSI: $\sin^2 \theta_{23} = 0.5$, $\Delta m^2_{32}=2.62\times 10^{-3}$~eV$^2$, $\epsilon_{\tau\tau}=0.6$ and $\epsilon_{e\tau}=1.2$.  The normal hierarchy and $\sin^2 \theta_{13}=0.030$ is assumed. For the other parameter values, see the text. The shaded bands illustrate the size of the 1$\sigma$ systematic uncertainties. The solid (dashed) curves in the right panel correspond to the NO$\nu$A (T2K) experiment.}
	\label{fig:compare}
\end{figure*}

In Fig.~\ref{fig:chiett}, we display the confidence level at which maximal mixing is excluded as a function of $\epsilon_{\tau\tau}$ after marginalizing over $|\epsilon_{e\tau}|< 1.2$ 
for both the normal and inverted hierarchy, and for two possible values of $\theta_{13}$. 
We choose $\sin^2 \theta_{13}=0.022$, which is a weighted average of recent reactor neutrino results, and $\sin^2 \theta_{13}=0.030$, which is at the edge of the 3$\sigma$ range obtained in Ref.~\cite{Gonzalez-Garcia:2013usa}. We checked that varying $\theta_{13}$ within the 3$\sigma$ allowed range has very little effect on the exclusion of maximal mixing in the SM analysis. The exclusion weakens significantly for some large values of $|\epsilon_{\tau\tau}|$. For the normal hierarchy, an NSI scenario with $\epsilon_{\tau\tau}=0.6$ and $\epsilon_{e\tau}=1.2$ is perfectly consistent with $\theta_{23}=\pi/4$. (Incidentally, $\chi^2_\text{min}(0.6,1.2) = 4.39$ which represents a very good fit to the NO$\nu$A data.)
Inserting $\epsilon_{\tau\tau}=0.6$, $\epsilon_{e\tau}=1.2$, $\epsilon_{\mu\mu}=\epsilon_{\mu\tau}=0$, $\theta_{23}=\pi/4$, $\sin^2 \theta_{13}=0.030$ and $\hat{A}(E_\nu=1.625~{\rm{GeV}})\approx 0.14$ into Eqs.~(\ref{eq:delta23}--\ref{eq:amp3}) gives $\sin^2 \theta=0.41$, which is close to the best-fit obtained by
NO$\nu$A.

In Fig.~\ref{fig:chiett}, the lack of symmetry of the curves about $\epsilon_{\tau\tau}=0$ can be understood from Eqs.~(\ref{eq:delta23}) and~(\ref{eq:amp3}). For $\epsilon_{\tau\tau}>0$, we have $\alpha_{23}<0$ for the normal hierarchy. 
 The sign of $\alpha_{23}$ has a negligible effect on $\sin 2(\theta_{23}+\alpha_{23})$ $(\approx \sin 2\theta_{23}\cos 2\alpha_{23}$ for $\theta_{23} \approx \pi/4$).
 However, the negative sign of $\alpha_{23}$ can significantly reduce 
 $\sin (\theta_{23}+\alpha_{23})$. This means that for a fixed magnitude of $\epsilon_{\tau\tau}$,  $\epsilon_{\tau\tau}>0$ will lead to a smaller value of the oscillation amplitude $\sin^2 2\theta$ than for $\epsilon_{\tau\tau}<0$. Since the experimentally preferred values have $\sin^2 2\theta\approx 0.95$, for the normal hierarchy the positive branch of $\epsilon_{\tau\tau}$ will reach the minimum of $\sqrt{\Delta \chi^2(\epsilon_{\tau\tau})}$ for smaller values of $|\epsilon_{\tau\tau}|$ than the negative branch. For the inverted hierarchy, $\hat{A}<0$, so the effect is reversed.

In Fig.~\ref{fig:compare} we plot the event distributions and survival probabilities for three different scenarios with a normal mass hierarchy. The SM(a) scenario has parameters close to the best-fit values from the T2K experiment~\cite{Abe:2014ugx} with maximal mixing. The SM(b) scenario corresponds to the best-fit values from the recent NO$\nu$A measurement~\cite{NOvAdata}. For the NSI scenario, we choose $\theta_{23}=\pi/4$ with $\epsilon_{\tau\tau}=0.6$ and $\epsilon_{e\tau}=1.2$ and all other NSI parameters set to zero. In both panels we see that the NSI scenario with maximal mixing is substantially similar to the SM(b) scenario with nonmaximal mixing. The NO$\nu$A measurement of nonmaximal mixing in the standard scenario could be interpreted as a hint for NSI with maximal mixing. The T2K curves in the right panel are almost overlapping for the NSI and SM(a) scenarios because T2K has a relatively short baseline. The small difference between them is due to the different values of $\Delta m^2_{32}$.


It is noteworthy that with three years of data in the neutrino mode and three years in the antineutrino mode, NO$\nu$A will differentiate between the NSI scenario with maximal mixing depicted in Fig.~\ref{fig:compare} and the SM scenario with nonmaximal mixing at about the $3\sigma$ CL. 

{\bf Summary.} We analyzed the recent NO$\nu$A $\nu_\mu$ disappearance data in the framework of nonstandard neutrino interactions. We find that if the NSI parameters $|\epsilon_{e\tau}|$ and $|\epsilon_{\tau\tau}|$ are ${\cal O}(1)$, the recent NO$\nu$A data are consistent with $\theta_{23}=\pi/4$, a value preferred by the T2K data. 
${\cal O}(1)$ values for these NSI parameters have a negligible effect on the T2K measurement. This means that the value of $\theta_{23}$ measured by T2K is close to the vacuum value, while the nonmaximal mixing detected by NO$\nu$A could be a hint of matter NSI. 

We consider our study to be a proof of principle that demonstrates that if this anomaly blossoms into something more significant, then currently running experiments may lead us to NSI, and that we may not have to wait more than a decade for experiments like DUNE and T2HK to discover the existence of NSI~\cite{Liao:2016hsa,Col}. In fact, the drastically altered mission of future long-baseline experiments would be to corroborate this discovery.


\vspace{0.1 in}
{\it Acknowledgments.} This research was supported by the U.S. DOE under Grant No. DE-SC0010504.


\vskip1cm


\end{document}